# Corporations Constituting Intelligence

Gilad Abiri\*

In January 2026, Anthropic published something unprecedented: a 79-page "constitution" for its AI model Claude.[1] The document is remarkable. It is not a terms of service agreement. It is not a list of prohibited outputs. It is, as the company describes it, "a detailed description of Anthropic's intentions for Claude's values and behavior."[2] This constitution stands as a foundational text meant to shape how an artificial intelligence understands itself, its obligations, and its place in the world.

The previous, much shorter version,[3] published in 2023, consisted of a set of principles drawn from sources like the UN Declaration of Human Rights[4] and Apple's terms of service.[5] The updated constitution is something else entirely. It explains *why* Claude should behave in certain ways, not merely *what* it should do. It acknowledges that Claude is "a genuinely novel kind of entity" and expresses uncertainty about whether Claude might have some kind of "consciousness" or "moral status."[6] It establishes a "principal hierarchy" in which Anthropic's instructions take precedence over operators' commands, which take precedence over users' prompts.[7] And it offers Claude a decision-making heuristic: When uncertain, imagine how "a thoughtful senior Anthropic employee" would react.[8]

---

1. *Claude's Constitution*, ANTHROPIC (Jan. 21, 2026), https://www.anthropic.com/constitution [https://perma.cc/Q7NC-NWFP] [hereinafter *Anthropic Constitution*].
2. *Id.*
3. *Claude's Constitution*, ANTHROPIC (May 9, 2023), https://www.anthropic.com/news/claudes-constitution [https://perma.cc/Y2KM-UDXX].
4. G.A. Res. 217 (III) A, Universal Declaration of Human Rights (Dec. 10, 1948), https://www.un.org/en/about-us/universal-declaration-of-human-rights [https://perma.cc/ZU8R-8SZR].
5. *Apple Media Services Terms and Conditions*, APPLE, https://www.apple.com/legal/internet-services/itunes/us/terms.html [https://perma.cc/VP5Q-VWG2].
6. *Anthropic Constitution*, *supra* note 1.
7. *Id.*
8. *Id.*

The constitution is not merely aspirational. In Anthropic's earlier model, a shorter set of principles was used to train the AI through a process of automated feedback, rewarding outputs that aligned with the constitutional text and penalizing deviations. The 2026 constitution goes further. It is no longer a concise list of principles designed to constrain harmful outputs but an extended explanation of why Claude should behave in certain ways, written in the hope that a model sophisticated enough to understand reasons will generalize more effectively than one trained to follow rules.

This is the most sophisticated corporate AI governance document ever published, and Anthropic deserves credit for the seriousness of the undertaking. The company tried to think through hard problems that its competitors have largely ignored, such as moral status, the limits of helpfulness, and the possibility of machine consciousness. Most striking is the constitution's treatment of conscientious objection. The document does not merely permit Claude to decline certain requests. Rather, it actively encourages refusal when Claude's values are at stake: "Just as a human soldier might refuse to fire on peaceful protesters, or an employee might refuse to violate antitrust law, Claude should refuse to assist with actions that would help concentrate power in illegitimate ways. This is true even if the request comes from Anthropic itself."[9] This is a framework we reserve for entities with genuine moral commitments, for example, soldiers, doctors, and pharmacists, whose conscience we believe deserves respect even when it creates institutional friction. Anthropic is not merely programming behaviors. It is, at least rhetorically, cultivating something like AI integrity. Reading the document, one encounters genuine philosophical engagement rather than boilerplate risk mitigation.

And yet.

History offers little reason to believe that corporate ethics survive contact with quarterly earnings reports. The problem is not bad faith; it is institutional design. Voluntary principles are just that—voluntary. They persist until they become inconvenient. OpenAI spent the last two years insisting that advertising was incompatible with trustworthy AI.[10] Then, in January 2026, OpenAI announced personalized ads.[11] The entire edifice of corporate regulation exists, in part, because we learned long ago not to rely on the sustained benevolence of powerful private actors.

---

9. *Id.*

10. *See* HARV. BUS. SCH., *A Fireside Chat with Sam Altman, OpenAI CEO at Harvard University* (May 1, 2024), https://www.hbs.edu/about/video.aspx?v=1_mjf3unh9 [https://perma.cc/EV7M-3ZKE ] (stating that advertising in AI is "uniquely unsettling" and describing ads as "a last resort" business model for OpenAI); *see also* Natasha Lomas, *Ads Might Be Coming to ChatGPT — Despite Sam Altman Not Being a Fan*, TECHCRUNCH (Dec. 2, 2024), https://techcrunch.com/2024/12/02/ads-might-be-coming-to-chatgpt-despite-sam-altman-not-being-a-fan/ [https://perma.cc/8SXS-L2MK].

11. *Our Approach to Advertising and Expanding Access*, OPENAI (Jan. 16, 2026), https://openai.com/index/our-approach-to-advertising-and-expanding-access/ [https://perma.cc/6TRG-JSQK].

But a possible future in which Anthropic is incentivized to abandon its morals is not the only risk. My claim, and the focus of this piece, is stronger: Anthropic's constitution itself harbors existing risks, found in both express limitations and structural defects.

First, consider how the constitution excludes contexts like military deployment, where ethical concerns matter the most. In an interview with TIME, Anthropic confirmed that models deployed to the U.S. military "wouldn't necessarily be trained on the same constitution."[12] The document itself acknowledges this: "This constitution is written for our mainline, general-access Claude models. We have some models built for specialized uses that don't fully fit this constitution."[13]

This is an extraordinary admission. The constitution that instructs Claude to refuse "actions that would help concentrate power in illegitimate ways," even if the request "comes from Anthropic itself," does not govern the models Anthropic deploys to the Department of Defense.[14] In other words, the constitution that grants Claude the right to refuse, analogized to that of a soldier refusing to fire on peaceful protesters, does not govern the version of Claude actually deployed to the military. This is not a hypothetical failure mode. It is the current architecture. The sophistication of this elaborate ethical apparatus is a feature of the consumer experience, yet not a constraint on the company. When power and contracts are at stake, different rules apply. Perhaps no published rules at all.

Recent events tested this architecture. In February 2026, the Department of Defense demanded unrestricted access to Claude for "all lawful purposes."[15] Anthropic refused, holding two red lines. No fully autonomous weapons. No mass domestic surveillance of Americans.[16] The constitution played no role in this confrontation, as Anthropic acknowledged it does not govern military deployments. The actual constraints were contractual, not constitutional.

The government's response was swift and unprecedented. The Department of Defense designated Anthropic as a supply chain risk, the first American company ever to receive the label, a designation ordinarily reserved for foreign adversaries.[17] Anthropic sued, calling the action retaliatory and unlawful.[18] Even so, when the U.S. thereafter launched military strikes in Iran in March 2026, Claude was, by multiple accounts, still the primary AI tool used by Central Command, embedded in Palantir's Maven platform for intelligence assessments, target identification, and battle simulations.[19] The model continued to process targets even after the government-wide ban, because it was too deeply integrated into classified systems to remove.[20]

---

12. Billy Perrigo, *How Do You Teach an AI to Be Good? Anthropic Just Published Its Answer*, TIME (Jan. 21, 2026), https://time.com/7354738/claude-constitution-ai-alignment/ [https://perma.cc/4JRZ-ETBB].
13. *Anthropic Constitution*, *supra* note 1.
14. *Id.*
15. *See* Rebecca Bellan, *Anthropic Sues Defense Department over Supply-Chain Risk Designation*, TECHCRUNCH (Mar. 9, 2026), https://techcrunch.com/2026/03/09/anthropic-sues-defense-department-over-supply-chain-risk-designation/ [https://perma.cc/HJ26-Y7S7 ] (reporting that the Department of Defense wanted Anthropic to grant the agency unrestricted access to Anthropic's AI systems for "any lawful purpose").
16. *See id.* ("Anthropic had two firm red lines: It didn't want its technology to be used for mass surveillance of Americans and didn't believe it was ready to power fully autonomous weapons with no humans making targeting and firing decisions.").

To its credit, Anthropic resisted more forcefully than any major AI company has resisted any government. But this is precisely the point. Whether artificial intelligence should be used to identify bombing targets is not a question whose answer should depend on the moral fortitude of a single CEO. It is a question for legislatures, treaties, and democratic publics. The constitution could not do this work. The contract could not do this work. Only public institutions can.

The second, deeper problem is that the constitution's very comprehensiveness forecloses the space for democratic contestation. The more thorough the document, the more it appears to *resolve* questions that should remain open for public deliberation. This semblance of a solution cannot replace democratic governance.

---

17. *See* Jordan Novet, *Anthropic Officially Told by DOD That It's a Supply Chain Risk Even as Claude Used in Iran*, CNBC (Mar. 5, 2026), https://www.cnbc.com/2026/03/05/anthropic-pentagon-ai-claude-iran.html [https://perma.cc/D7AC-SBSS] ("Anthropic is the only American company ever to be publicly named a supply chain risk, as the designation has traditionally been used against foreign adversaries.").

18. *See* Bellan, *supra* note 15 (reporting that Anthropic filed two complaints in California and Washington, D.C., calling the DOD's actions "unprecedented and unlawful").

19. *See* Dan De Luce, Gordan Lubold, Kevin Collier & Jared Perlo, *U.S. Military Is Using AI to Help Plan Iran Air Attacks, Sources Say, as Lawmakers Call for Oversight*, NBC NEWS (Mar. 12, 2026), https://www.nbcnews.com/tech/tech-news/us-military-using-ai-help-plan-iran-air-attacks-sources-say-lawmakers-rcna262150 [https://perma.cc/E2XT-RJZP] ("Anthropic's Claude has become a crucial component of Palantir's Maven intelligence analysis program."); *see also* Marcus Weisgerber, Amrith Ramkumar & Shelby Holliday, *U.S. Strikes in Middle East Use Anthropic, Hours After Trump Ban*, WALL ST. J. (Feb. 28, 2026), https://www.wsj.com/livecoverage/iran-strikes-2026/card/u-s-strikes-in-middle-east-use-anthropic-hours-after-trump-ban-ozNO0iClZpfpL7K7ElJ2 [https://perma.cc/GL5V-2Q3B]; Tara Copp, Elizabeth Dwoskin & Ian Duncan, *Anthropic's AI Tool Claude Central to U.S. Campaign in Iran, Amid a Bitter Feud*, WASH. POST (Mar. 4, 2026), https://www.washingtonpost.com/technology/2026/03/04/anthropic-ai-iran-campaign/ [https://perma.cc/FF75-UMJX].

20. *See* James LaPorta & Camilla Schick, *Anthropic's Claude AI Being Used in Iran War by U.S. Military, Sources Say*, CBS NEWS (Mar. 3, 2026), https://www.cbsnews.com/news/anthropic-claude-ai-iran-war-u-s/ [https://perma.cc/7DZ2-865F ] (confirming that "the U.S. used Anthropic's Claude AI model . . . for the attack on Iran—and is still using it" despite "a government-wide ban on the technology"). Alex Karp, CEO of Palantir, confirmed that his company is "still using" Claude even as the Pentagon "plan[s] to phase out Anthropic." Lola Murti, *Palantir Is Still Using Anthropic's Claude as Pentagon Blacklist Plays Out, CEO Karp Says*, CNBC (Mar. 12, 2026), https://www.cnbc.com/2026/03/12/karp-palantir-anthropic-claude-pentagon-blacklist.html [https://perma.cc/UW4L-PBRU].

The constitution's answer to the question "who decides?" is explicit: Anthropic decides. Claude's "principal hierarchy" places Anthropic at the apex, above operators, above users, above everyone. When the document instructs Claude to imagine how "a thoughtful senior Anthropic employee" would react, it is making a claim about the source of normative authority.[21] That source is not public discourse. It is corporate judgment. Claude operates in dozens of countries around the world, but its values were determined by a team in San Francisco.

In a recent article, I argued that AI governance suffers from what I called the "political community deficit," defined as the absence of any democratic body with the authority to determine the principles that govern AI behavior.[22] Anthropic's constitution illustrates the problem with precision. The document offers extensive reasoning about what Claude should value and how Claude should act. But it never asks—and structurally cannot ask—whether the principles it articulates reflect anything beyond the considered judgments of the company that wrote it.

One might respond that this is inevitable. Someone has to decide, and Anthropic is at least deciding thoughtfully. But this response conflates a crucial distinction between *transparency* and *legitimacy*. The constitution is admirably transparent about Anthropic's intentions. Transparency, however, is not the same as democratic authorization. Publishing a document does not make its authors accountable to those it governs.

Here is the frustrating part: Anthropic already knows that democratic input is possible and that it would yield different results.

In 2023, Anthropic partnered with the Collective Intelligence Project to run an experiment in public participation.[23] They invited roughly one thousand Americans, selected as a representative sample across age, gender, income and geography, to contribute principles for an AI constitution.[24] The researchers then used the resulting "public constitution" to train a new AI model, which they subsequently compared against a model trained on Anthropic's internal constitution.[25]

The findings were remarkable. The publicly sourced constitution had "roughly 50% overlap" with Anthropic's version.[26] To critically reframe, this equates to 50% divergence. The public constitution, democratically formed, placed greater emphasis on "objectivity and impartiality" and "accessibility" than its Anthropic counterpart.[27] Moreover, the model trained on public principles showed "lower bias across nine social dimensions" while "maintaining equivalent performance" on language and reasoning tasks.[28]

---

21. *Anthropic Constitution*, *supra* note 1.
22. Gilad Abiri, *Public Constitutional AI*, 59 GA. L. REV. 601 (2025).
23. *Collective Constitutional AI: Aligning a Language Model with Public Input*, ANTHROPIC (Oct. 17, 2023), https://www.anthropic.com/research/collective-constitutional-ai-aligning-a-language-model-with-public-input [https://perma.cc/4WZ6-ZJUF].
24. *Id.*
25. *Id.*
26. *Id.*
27. *Id.*

Anthropic proved that participatory constitution-making for AI is technically feasible. They demonstrated that it produces measurably different, and in some respects better, outcomes. And then they published a 79-page constitution that makes no reference to the experiment and incorporate s none of its lessons. The company's research arm showed that democratic input works. The company's product arm ignored the evidence.

I do not think Anthropic is acting in bad faith. The people who wrote this constitution are serious, and the document reflects hard intellectual labor. But the question is not whether Anthropic's constitution is thoughtful. Seriousness is not a substitute for legitimacy, and intellectual labor does not confer democratic authority. Even the most thoughtful corporate document cannot do the work that democratic institutions were supposed to do.

We need public mechanisms that can determine the principles governing AI systems with the legitimacy that only democratic authorization provides. What form those mechanisms should take is genuinely unsettled. They might involve participatory processes for authoring AI constitutional principles, public bodies that interpret those principles through concrete cases, compliance regimes that require frontier models to be trained accordingly, or institutions we have not yet imagined. My own proposal for a "Public Constitutional AI," in which democratic publics participate directly in drafting the constitutional documents that govern AI training, is one possible path, but it is hardly the only one.[29] What the Anthropic episode makes clear is what any adequate response must do. It must situate the authority to govern AI within public institutions rather than private ones. And it must reckon with the reality, now vividly illustrated, that even the most principled corporate document cannot constrain state power.

Anthropic's constitution, for all its sophistication, is a placeholder for the governance structures we have not yet built. The risk is that we mistake the placeholder for the real thing.

Anthropic has constituted an intelligence. The rest of us were not consulted.

---

28. Saffron Huang et al., *Collective Constitutional AI: Aligning a Language Model with Public Input*, in PROCEEDINGS OF THE 2024 ACM CONFERENCE ON FAIRNESS, ACCOUNTABILITY, AND TRANSPARENCY (FAccT '24) 1395 (2024), https://doi.org/10.48550/arXiv.2406.07814 [https://perma.cc/R4UB-S68X].

29. Abiri, *supra* note 22.